\documentclass[epj,nopacs]{svjour}
\usepackage{amssymb,amsmath}
\usepackage{graphicx,color}

\newcommand{\dd}{\mathrm{d}}
\newcommand{\ii}{\mathrm{i}}
\DeclareMathOperator{\im}{Im}
\newcommand{\erw}[1]{\ensuremath {\left \langle {#1} \right \rangle}}

\title{T-matrix approach to heavy quark diffusion in the QGP}

\author{H. van Hees \inst{1} \and M. Mannarelli\inst{2} \and
  V. Greco\inst{3} \and R. Rapp\inst{4}}

\institute{Institut f{\"u}r Theoretische Physik,
   Justus-Liebig-Universit{\"a}t Giessen, Heinrich-Buff-Ring 16, D-35392
  Giessen, Germany \and
Instituto de Ciencias del Espacio
    (IEEC/CSIC), E-08193 Bellaterra (Barcelona), Spain \and
Dipartimento di Fisica e Astronomia, Via S. Sofia 64,
  I-95125 Catania, Italy \and
Cyclotron Institute and Physics Department,
       Texas A\&M University, College Station, Texas 77843-3366, U.S.A.
}

\date{August 27, 2008} 

\abstract{We assess transport properties of heavy quarks in the
  Quark-Gluon Plasma (QGP) using static heavy-quark (HQ) potentials from
  lattice-QCD calculations in a Brueckner many-body $T$-matrix approach
  to evaluate elastic heavy-quark-light-quark scattering amplitudes. In
  the attractive meson and diquark channels resonance states are formed
  for temperatures up to $\sim$$1.5 T_c$, increasing pertinent drag and
  diffusion coefficients for heavy-quark rescattering in the QGP beyond
  the expectations from perturbative-QCD calculations. We use these
  transport coefficients, complemented with perturbative elastic HQ
  gluon scattering, in a relativistic Langevin simulation to obtain HQ
  $p_t$ distributions and elliptic flow ($v_2$) under conditions
  relevant for the hot and dense medium created in ultrarelativistic
  heavy-ion collisions. The heavy quarks are hadronized to open-charm
  and -bottom mesons within a combined quark-coalescence fragmentation
  scheme. The resulting single-electron spectra from their semileptonic
  decays are confronted with recent data on ``non-photonic electrons''
  in $200 \; A\mathrm{GeV}$ Au-Au collisions at the Relativistic
  Heavy-Ion Collider (RHIC).}

\begin{document}
\maketitle

\section{Introduction}
\label{sec-intro}

One of the most interesting questions in high-energy nuclear physics is
that about the properties of the hot and dense medium created in
ultra-relativistic heavy-ion collisions. Finite-temperature lattice-QCD
(lQCD) calculations of strong\-ly-in\-ter\-ac\-ting matter predict a
phase transition from hadronic matter to a quark-gluon plasma (QGP) at a
critical temperature, $T_c \simeq 180 \;
\mathrm{MeV}$~\cite{Karsch:2007vw}. In the recent years the experimental
program at the Relativistic Heavy-Ion collider has resulted in
convincing evidence for the formation of such a hot and dense partonic
state~\cite{Arsene:2004fa,Back:2004je,Adams:2005dq,Adcox:2004mh}.

The heavy charm and bottom quarks are particularly valuable probes for
the properties of this medium since they are created in the primordial
hard collisions of the nucleons within the colliding nuclei. Thus, they
form a rather well defined initial state and interact with the hot and
dense fireball during its entire evolution. Recently, measurements of
the transverse-momentum distributions of ``non-photonic single
electrons'' ($e^{\pm}$), which originate mainly from the semi-leptonic
decays of open-charm and -bottom mesons, in $200 \; A\mathrm{GeV}$ Au-Au
collisions at RHIC have found a surprisingly large suppression at high
transverse momenta ($p_t$) (i.e., a small nuclear modification factor,
$R_{AA}$) and a large elliptic-flow parameter, $v_2$. Both findings
indicate that during the lifetime of the hot and dense fireball heavy
quarks come close to thermal equilibrium with the
medium~\cite{Adler:2005xv,Abelev:2006db,Adare:2006nq}.

The theoretical challenge is to understand the corresponding
thermalization times of heavy quarks from the underlying microscopic
scattering processes with the constituents of the QGP, in particular how
the heavy quarks, despite their large masses, $m_Q \gg T_c$, become part
of the collective flow of the fireball. In calculations of the pertinent
transport coefficients from perturbative QCD (pQCD), based on
gluon-brems\-strah\-lung energy loss, including elastic HQ scattering,
one has to artificially tune the coupling strength beyond the
applicability range of perturbation
theory~\cite{Armesto:2005mz,Wicks:2005gt}. It has also been shown that
the convergence of the perturbative series for the HQ diffusion
coefficient is quite poor~\cite{CaronHuot:2007gq}. Thus,
non-per\-tur\-ba\-tive approaches have to be used to explain the strong
HQ couplings necessary. One suggested mechanism is the formation of $D$-
and $B$-meson resonance excitations in the deconfined phase of QCD
matter~\cite{vanHees:2004gq,vanHees:2005wb} which has lead to a quite
satisfactory description of the $e^{\pm}$ data at RHIC.

This paper is organized as follows: In Sec.~\ref{sec-t-matrix} we use HQ
static potentials from lattice-QCD calculations at finite temperature in
a many-body Brueckner $T$-matrix approach to calculate elastic HQ
light-quark scattering-ma\-trix elements in the medium
\cite{Mannarelli:2005pz,vanHees:2007me}. We show that after inclusion of
a complete set of color channels, taking into account $l=0,1$ states in
the partial-wave expansion of the $T$-matrix, the resonance states,
conjectured in the earlier approaches, are confirmed by these
interactions, which are in principle free of tunable parameters. The
resulting elastic-scattering amplitudes are used in
Sec.~\ref{sec-langevin} to calculate drag and diffusion coefficients for
a Fokker-Planck
equation~\cite{Svetitsky:1987gq,vanHees:2004gq,Moore:2004tg}, describing
the rescattering of the heavy quarks within the hot and dense sQGP
fireball. In the next step we employ a relativistic Langevin simulation
to find the corresponding HQ $p_t$ distributions, using a
thermal-fireball parameterization, including elliptic flow for
non-central heavy-ion collisions. To confront these spectra with the
$e^{\pm}$ data from the PHENIX and STAR collaborations at RHIC, in
Sec.~\ref{sec-observables} we use a combined quark-coalescence and
fragmentation model to hadronize the heavy quarks to $D$ and $B$ mesons
which then are decayed semi-leptonically leading to the final $e^{\pm}$
spectra which can be directly confronted with recent data on nonphotonic
single electrons in $200 \; A \text{GeV}$ Au-Au collisions at RHIC. The
paper closes with brief conclusions and an outlook
(Sec.~\ref{sec-conclusions}).

\section{HQ scattering in the QGP}
\label{sec-t-matrix}

In this Sec. we calculate in-medium matrix elements for elastic
scattering of heavy quarks ($Q=c,b$) with light quarks $q=u,d,s$ in a
Brueckner-like many-body approach, assuming that a static heavy-quark
light-quark potential, $V(r)$, can be employed as the interaction
kernel. Such a model has been used in the vacuum to successfully
describe $D$-meson spectra and
decays~\cite{Godfrey:1985xj,Avila:1994vi}. Further, we assume that the
effective in-medium potential can be extracted from finite-temperature
lQCD calculations of the color-singlet free energy
$F_1(r,T)$~\cite{Kaczmarek:2003dp,Kaczmarek:2005gi} for a static
$\bar{Q} Q$ pair as the internal potential energy by the usual
thermodynamic
relation~\cite{Mannarelli:2005pz,Shuryak:2004tx,Wong:2004zr,Cabrera:2006wh},
\begin{equation}
\label{F-to-U}
U_1(r,T)=F_1(r,T)-T \frac{\partial F_1(r,T)}{\partial T}.
\end{equation}
For application as a scattering kernel in a $T$-matrix equation, the
potential as to vanish for $r \rightarrow \infty$. Thus we choose the
accordingly subtracted internal potential energy,
\begin{equation}
\label{U-to-V}
V_1(r,T)=U_1(r,T)-U_1(r \rightarrow \infty,T).
\end{equation}
In lQCD simulations one finds that $U_1(r \rightarrow \infty,T)$ is a
decreasing function with temperature which could be associated as a
contribution to the in-medium HQ mass, $m_Q(T)=m_0+U_1(r \rightarrow
\infty,T)/2$ where $m_0$ denotes the bare mass. However, this leads to
problems since close to $T_c$ the asymptotic value, $U_1(r \rightarrow
\infty,T)$, develops a pronounced peak structure. Thus, in this
calculation, we assume constant effective HQ masses, $m_c=1.5 \;
\mathrm{GeV}$ and $m_b=4.5 \; \mathrm{GeV}$.

We also consider the complete set of color channels for the $Q \bar{q}$
(singlet and octet) and $Q q$ (anti-triplet and sextet) systems, using
Casimir scaling as in leading-order pQCD,
\begin{equation}
\label{casimir-scaling}
V_{8}=-\frac{1}{8} V_1, \quad V_{\bar{3}}=\frac{1}{2} V_1, \quad
V_6=-\frac{1}{4} V_1,
\end{equation}
which is also justified by recent lQCD calculations of the finite-$T$ HQ
free energy~\cite{Nakamura:2005hk,Doring:2007uh}.

This approach is in principle parameter free in the choice of the
interactions, since their strength is taken from first-principle lQCD
simulations. However, there are considerable uncertainties in the
potentials (a) between different lattice calculations and (b) in the
extraction and parameterization of the corresponding free energies,
particularly their temperature dependence needed to subtract the entropy
term in Eq.~(\ref{F-to-U}). In addition, the very notion of an
``in-medium potential'' is not a unique concept~\cite{Brambilla:2008cx},
and its identification with the internal potential energy may be seen as
an upper limit in interaction strength. We use three different
parameterizations of $F_1$
\cite{Wong:2004zr,Shuryak:2004tx,Mannarelli:2005pz}:
\begin{description}
\item[{\qquad [Wo]}] of quenched lQCD~\cite{Kaczmarek:2003dp},
\item[{\qquad [SZ]}] of two-flavor
  lQCD~\cite{Kaczmarek:2003ph}, and
\item[{\qquad [MR]}] of three-flavor
  lQCD~\cite{Petreczky:2004pz}.
\end{description}
The resulting potentials from [Wo] and [SZ] are comparable to a
numerical extraction from three-flavor lQCD~\cite{Petreczky:2004pz},
while that from [MR] is deeper than the other two for $T \lesssim 1.6
T_c$, but falls off faster at higher temperatures. The resulting
uncertainty in the transport coefficients (see Sec.~\ref{sec-langevin})
amounts to up to $40 \%$.
\begin{figure}
\centerline{\includegraphics[width=0.9\linewidth]{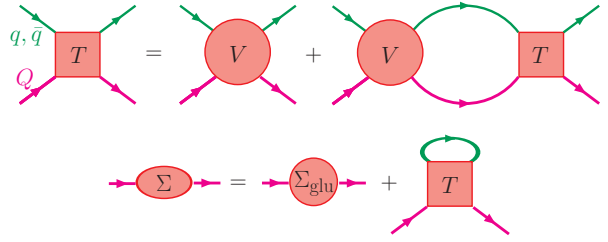}}
\caption{(Color online) Diagrammatic representation of the Brueckner
  many-body scheme for the coupled system of the $T$-matrix based on the
  lQCD static internal potential energy as the interaction kernel and
  the HQ self-energy.}
\label{fig.brueckner}
\end{figure}
To define the Brueckner-type many-body scheme the four-di\-men\-sio\-nal
(4D) Bethe-Salpeter (BS) ladder approximation, symbolized in
diagrammatical form by the upper panel of Fig.~\ref{fig.brueckner}, has
to be reduced to a 3D Lippmann-Schwinger (LS) equation, neglecting
antiparticle components in the quark propagators, in order to implement
the static potential from lQCD
via~Eqs.~(\ref{F-to-U}-\ref{casimir-scaling}). After this reduction the
LS equation in the color channel, $a \in \{1,\bar{3},6,8\}$
reads~\cite{Mannarelli:2005pz}\footnote{Here and in the following all
  vertex and Green's functions are understood as the retarded real-time
  quantities which can be derived as analytic continuations of the
  corresponding imaginary-time (Matsubara) quantities of thermal quantum
  field theory.}
\begin{equation}
\begin{split}
\label{LS}
T_a(E;\vec{q}',\vec{q}) = &V_a(\vec{q}',\vec{q})-\int \frac{\dd^3
  \vec{k}}{(2 \pi)^3} V_q(\vec{q}',\vec{k}) G_{qQ}(E;k) \\
& \times T_a(E;\vec{k},\vec{q})[1-f_F(\omega_k^q)-f_F(\omega_k^Q)]
\end{split}
\end{equation}
with the Fourier-transformed potentials,
\begin{equation}
\label{v-four}
V_a(\vec{q}',\vec{q})=\int \dd^3 \vec{r} V_a(r) \exp[\ii(\vec{q}-\vec{q}')\vec{r}].
\end{equation}
Further, $E$, $\vec{q}$ and $\vec{q}'$ denote the energy and incoming
and outgoing momenta in the center-of-mass (CM) frame, respectively.
\begin{equation}
\label{fd-dist}
f_F=\frac{1}{\exp(\omega/T)+1}
\end{equation}
is the Fermi-Dirac distribution. The quark-dispersion relations are
determined in quasi-particle approximation by
\begin{equation}
\omega_k^{q,Q}=\sqrt{k^2+m_{q,Q}^2},
\end{equation}
where for simplification we do not solve the fully self-consistent
scheme in Fig.~\ref{fig.brueckner} but use a fixed mass of $m_q=0.25 \;
\text{GeV}$, $m_c=1.5 \; \text{GeV}$, and $m_b=4.5 \; \text{GeV}$ for
the light, charm, and bottom quarks, respectively. Finally, the
two-particle-$qQ$ propagator in~(\ref{LS}) is given in terms of the
Thompson-reduction scheme~\cite{Thompson:1970wt}
\begin{equation}
\label{thompson-prop}
G_{qQ}(E;k)=\frac{1}{4} \frac{1}{E- (\omega_k^q +\ii
  \Sigma_I^q)-(\omega_k^Q +\ii \Sigma_I^Q)}
\end{equation}
with a quasi-particle width for both light and heavy quarks of
$-2\Sigma_{I}^{q,Q}=0.2 \; \text{GeV}$.

The solution of the LS equation (\ref{LS}) is simplified by using a
partial-wave expansion of the potential and $T$-matrix,
\begin{equation}
\begin{split}
\label{part-wave}
V_a(\vec{q}',\vec{q})=4 \pi \sum_{l} (2l+1) V_{a,l}(q',q) P_l[\cos
\angle(\vec{q},\vec{q}')], \\
T_a(E;\vec{q}',\vec{q})=4 \pi \sum_{l} (2l+1) T_{a,l}(E;q',q) P_l[\cos
\angle(\vec{q},\vec{q}')],
\end{split}
\end{equation}
which leads to the 1D LS equations,
\begin{equation}
\begin{split}
\label{LS-1D}
T_{a,l}(E;q',q)=V_{a,l}&(q',q) + \frac{2}{\pi} \int \dd k k^2
V_{a,l}(q',k) G_{Qq}(E;k) \\
&\times T_{a,l}(E;k,q)[1-f_F(\omega_k^Q)-f_F(\omega_k^q)],
\end{split}
\end{equation}
for the partial-wave components, $T_{a,l}$, of the $T$-matrix, which are
solved numerically with the matrix-inversion algorithm of Haftel and
Tabakin~\cite{Haftel1970}. We restrict ourselves to $S$ ($l=0$) and $P$
($l=1$) waves.
\begin{figure}
\centerline{\includegraphics[width=0.9\linewidth]{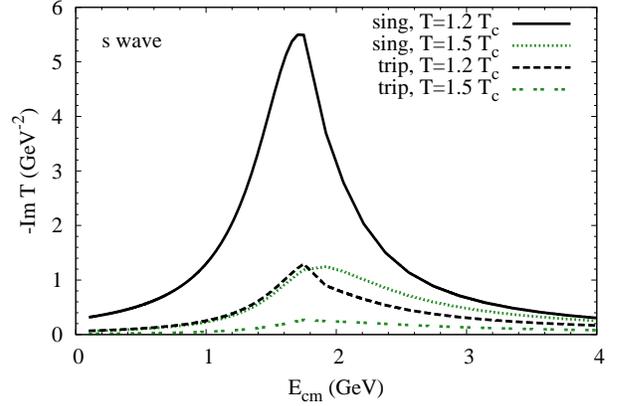}}
\caption{(Color online) Imaginary part of the $S$-wave in-medium $T$
  matrix for $c\bar{q}$ and $cq$ scattering in the color-singlet and
  -antitriplet channels based on the parameterization of the lQCD
  potential energy by [Wo].}
\label{fig.ImT}
\end{figure}
As can be seen from Fig.~\ref{fig.ImT}, in the dominating attractive
color-singlet $Q\bar{q}$ and color-antitriplet $Q q$ channels, close to
the critical temperature, $T_c \simeq 180 \; \text{MeV}$, resonance
states above the threshold, $E_{\text{thr}}=m_Q+m_q$ are formed, similar
as conjectured in~\cite{vanHees:2004gq,vanHees:2005wb}. However, in this
full in-medium scheme the resonances melt at higher temperatures $T
\gtrsim 1.7 T_c$ and $T \gtrsim 1.4 T_c$, respectively. As we shall see
in the next section, contrary to the expectation from perturbative
calculations, using the [Wo] parameterization of the potential, this
leads even to \emph{decreasing} transport coefficients with increasing
temperature, i.e., the decreasing interaction strength of the potential
overcompensates the higher density of the medium.

\begin{figure}
\centerline{\includegraphics[width=0.9\linewidth]{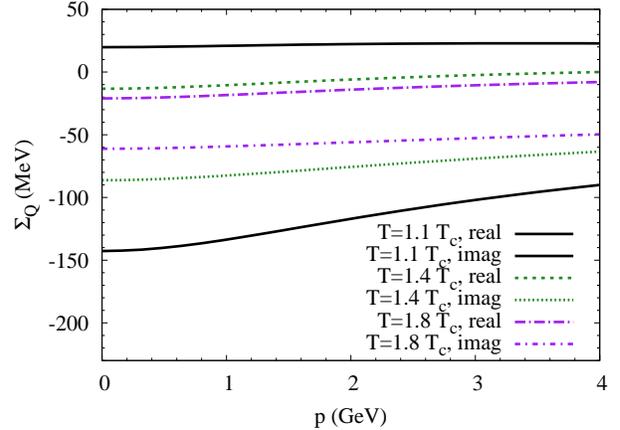}}
\caption{(Color online) Real and imaginary parts of the $c$-quark
  self-energy as a function of three-momentum at different
  temperatures.}
\label{fig.HQSE}
\end{figure}

The HQ self-energy, diagrammatically represented by the lower panel of
Fig.~\ref{fig.brueckner}, is then given by
\begin{equation}
\begin{split}
\label{self-en}
\Sigma_a^Q(\omega,p)=&\frac{d_{\text{SI}} d_a}{6} \int \frac{k^2 \dd k \dd
  x}{4\pi^2} [f_F(\omega_k)+f_B(\omega+\omega_k)] \\
& \times T_a(E;\vec{p},\vec{k}),
\end{split}
\end{equation}
where $d_{\text{SI}}=4(2l+1) N_f$ denotes the spin-isospin degeneracy of
the light quarks (which is already implicitly assumed in the light-heavy
quark interaction of our approach, which is in line with the free
$D$-meson spectrum \cite{Abe:2003zm}) in the $l^{\text{th}}$ partial
wave and $d_a$ the color degeneracy in the corresponding channel. Here
we assume an effective number of light-quark flavors, $N_f=2.5$, to
account for the smaller strange-quark density. For the charm quarks,
cf. Fig.~\ref{fig.HQSE} the calculation leads to an in-medium width of
$\Gamma_c=-2 \im \Sigma_c \simeq 200 \;\text{MeV}$, which justifies our
simplifying assumption in the $T$-matrix calculation above.

\section{HQ transport in the QGP}
\label{sec-langevin}

To evaluate the motion of the heavy quarks in the hot and dense
fireball, consisting of light quarks and gluons, we employ a Langevin
simulation of the Fokker-Planck equation,
\begin{equation}
\label{FP}
\frac{\partial f_Q}{\partial t}=\frac{\partial}{\partial p_i} (p_i \gamma
f_Q) + \frac{\partial^2}{p_i p_j} (B_{ij} f_Q).
\end{equation}
The drag or friction coefficient, $\gamma$, and diffusion coefficients,
\begin{equation}
\label{diff}
B_{ij}=B_0 \frac{p_i p_j}{p^2} + B_1 \left (1-\frac{p_i p_j}{p^2} \right ),
\end{equation}
are calculated from the invariant scat\-te\-ring-ma\-trix
elements~\cite{Svetitsky:1987gq}. Taking into account elastic scattering
of the heavy quark with a light quark or antiquark the latter given in
terms of the above calculated $T$-matrix by
\begin{equation}
\begin{split}
\label{Msq}
\sum |\mathcal{M}|^2=\frac{64\pi}{s^2} (s-m_q^2+m_Q^2)^2(s-m_Q^2+m_q^2)^2 \\
\times N_f\sum_{a} d_a (|T_{a,l=0}(s)|^2 +3 |T_{a,l=1} (s)
\cos(\theta_{\text{cm}})|^2).
\end{split}
\end{equation}
We define the averaging operator
\begin{equation}
\begin{split}
\label{erw}
\erw{X(\vec{p}{\,})} &= \frac{1}{2 E_p} \int \frac{\dd^3 \vec{q}}{(2
  \pi)^3 2E_q} \int \frac{\dd^3 \vec{q}{\,}'}{(2 \pi)^3 2E_{q'}} \\
& \quad \int \frac{\dd^3
  \vec{p}{\,}'}{(2\pi)^3 2E_{p'}} \frac{1}{\gamma_c} \sum |\mathcal{M}|^2 \\
& (2 \pi)^4 \delta^{(4)}(p+q-p'-q') f_q(\vec{q}) X(\vec{p}{\,}') \ ,
\end{split}
\end{equation}
for a heavy-quark observable, $X$, over the elastic scatterings per unit
time of the heavy quark with momentum $\vec{p}$ with a light quark of
momentum $\vec{q}$, changing their momenta to $\vec{p}'$ and $\vec{q}'$.
Here, $f_q$ is the (thermal) distribution of the light quarks in the
me\-dium. Then we can calculate the transport coefficients as
\begin{align}
\gamma(|\vec{p}|) &= \erw{1} -\frac{\erw{\vec{p} \cdot \vec{p}{\,}'}}{\vec{p}^2},
\label{A}
\\
B_0(|\vec{p}|) &= \frac{1}{4} \left [
  \erw{\vec{p}{\,}'{}^2}-\frac{\erw{(\vec{p} \cdot \vec{p}{\,}')^2}}{\vec{p}^2}
\right],
\label{B0}
\\
B_1(|\vec{p}|) &= \frac{1}{2} \left[
\frac{\erw{(\vec{p} \cdot \vec{p}{\,}')^2}}{\vec{p}^2} 
-2 \erw{\vec{p}{\,}' \cdot \vec{p}} + \vec{p}^2 \erw{1} \right]. 
\label{B1}
\end{align}
However, it turns out that, in order to guarantee the pro\-per equilibrium
limit of the heavy quarks with the me\-dium, we have to enforce Einstein's
fluctuation-dissipation relation for the longitudinal diffusion
coefficient, which is used here in its relativistic
form~\cite{Moore:2004tg,vanHees:2005wb},
\begin{equation}
\label{fluc-diss}
B_1=E T \gamma.
\end{equation}
The nonperturbative HQ light-quark scat\-te\-ring-ma\-trix ele\-ments are
supplemented by the corresponding perturbative elastic HQ
gluon-scattering ones~\cite{Combridge:1978kx}. The $t$-channel
singularity is regulated by a gluon-Debye screening mass of $m_g=g T$
with a strong coupling constant $g=\sqrt{4 \pi \alpha_s}$, using
$\alpha_s=0.4$.
\begin{figure}
\centerline{\includegraphics[width=0.9\linewidth]{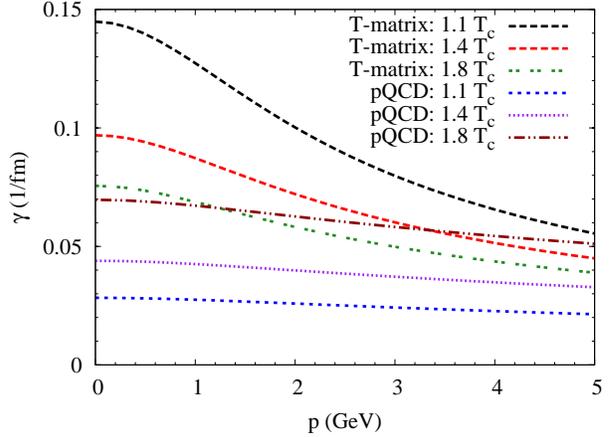}}
\caption{(Color online) The drag coefficient, $\gamma$, as a function of
  HQ momentum, calculated via (\ref{A}) with scattering-matrix elements
  from the non-perturbative $T$-matrix calculation (using the
  parameterization of the lQCD internal potential energies by [Wo])
  compared to a LO perturbative calculation based on matrix elements
  from~\cite{Combridge:1978kx}.}
\label{fig.frict-coeff}
\end{figure}
As shown in Fig.~\ref{fig.frict-coeff}, close to $T_c$ the equilibration
times of $\tau_{\text{eq}}=1/\gamma \simeq 7 \; \text{fm}/c$ for charm
quarks are a factor of $\sim 4$ larger than the values from a
corresponding pQCD calculation, reminiscent to the results based on the
model, assuming the survival of $D$-meson like resonance states above
$T_c$~\cite{vanHees:2004gq,vanHees:2005wb}. In contrast to this and
other calculations of the HQ transport coefficients, here the drag
coefficients \emph{decrease} with increasing temperature because of the
``melting'' of the dynamically generated resonances at increasing
temperatures due to the diminishing interaction strength from the lQCD
potentials.

To solve the Fokker-Planck equation~(\ref{FP}) under conditions of the
sQGP medium produced in heavy-ion collisions, we use an isentropically
expanding thermal fireball model, assuming an ideal-gas equation of
state of $N_f=2.5$ effective massless light-quark flavors and
gluons. The total entropy is fixed by particle multiplicities at
chemical freeze-out which we assume to occur at the critical
temperature, $T_c = 180 \; \text{MeV}$. For semi-central collisions the
fireball is chosen to be of elliptic-cylindrical shape with isobars
given by confocal ellipses with a perpendicular radial-flow field,
scaling linearly with the distance from the center as seen in
hydrodynamic calculations~\cite{Kolb:2000sd} to which also the (average)
radial flow velocity and ellipticity, $v_2$, is fixed. To compare to
``minimum-bias'' data on $e^{\pm}$ spectra in $200 \; A \text{GeV}$
Au-Au reactions at RHIC we simulate collisions with an impact parameter
of $b=7 \; \text{fm}$, implying an initial spatial eccentricity of about
$0.6$. Using a QGP-formation time of $0.33 \; \text{fm}/c$ leads to an
initial temperature of $340 \;\text{MeV}$. The evolution stops after the
fireball has undergone a mixed QGP-ha\-dro\-nic phase after about $5 \;
\text{fm}/c$, at which the radial velocity of the fireball has reached
about a radial velocity of about $v_{\perp}=0.5 c$ at the surface and an
ellipticity of $v_2 =5.5\%$.

Given this description of the medium, (\ref{FP}) is solved with help of
an equivalent relativistic Langevin simulation which is defined by the
stochastic equation of motion for a heavy quark at position, $\vec{x}$,
and momentum $\vec{p}$:
\begin{equation}
\label{langevin}
\delta \vec{x}=\frac{p}{E} \delta t, \quad \delta
\vec{p}=-\gamma(t,\vec{p}) \vec{p} \delta t + \delta
\vec{W}(t,\vec{p}+\delta \vec{p}), 
\end{equation}
where $\delta \vec{W}$ is a stochastic force distributed normally,
\begin{equation}
\label{rand-force}
P(\delta \vec{W}) \propto \exp \left [-\frac{(B^{-1})_{jk} \delta W^{j}
    \delta W^{k}}{4 \delta t} \right ],
\end{equation}
$B^{-1}$ denoting the inverse of the diffusion-coefficient matrix
(\ref{diff}). Note that in (\ref{langevin}) and (\ref{rand-force}) the
diffusion coefficients are to be evaluated at the updated momenta
$\vec{p}+\delta \vec{p}$. This H{\"a}nggi-Klimontovich realization of
the stochastic process together with the dis\-si\-pa\-tion-fluctuation
relation~(\ref{fluc-diss}) ensures the correct equilibrium limit in the
long-time regime $t \gg 1/\gamma$~\cite{Haenggi:2005}. After evaluation
of the time step (\ref{langevin}) the resulting momenta are Lorentz
boosted to the laboratory frame.

The initial condition for (\ref{langevin}) is given by the phase-space
distribution of the heavy quarks. The spatial distribution is determined
with a Glauber model for heavy-quark production. The initial $p_t$
spectrum is determined from data on $p$-$p$ and d-Au collisions at RHIC
as follows: The $c$-quark spectra are taken from a modified PYTHIA
calculation to fit $D$ and $D^*$ spectra in d-Au
collisions~\cite{Adams:2004fc}, assuming $\delta$-function
fragmentation. After decaying this spectrum to single $e^{\pm}$ they
saturate corresponding data from $p$-$p$- and
d-Au~\cite{Adler:2005xv,Tai:2004bf} collisions up to $p_t \simeq 3.5 \;
\text{GeV}$. The missing yield at higher $p_t$ is assumed to be filled
with the corresponding contributions from $B$ mesons, leading to a
cross-section ratio of $\sigma_{b\bar{b}}/\sigma_{c\bar{c}} \simeq 5 \cdot
10^{-3}$ and a crossing of the $c$- and $b$-decay electron spectra at
$p_t \simeq 5\;\text{GeV}$.
\begin{figure}
\centerline{\includegraphics[width=0.9\linewidth]{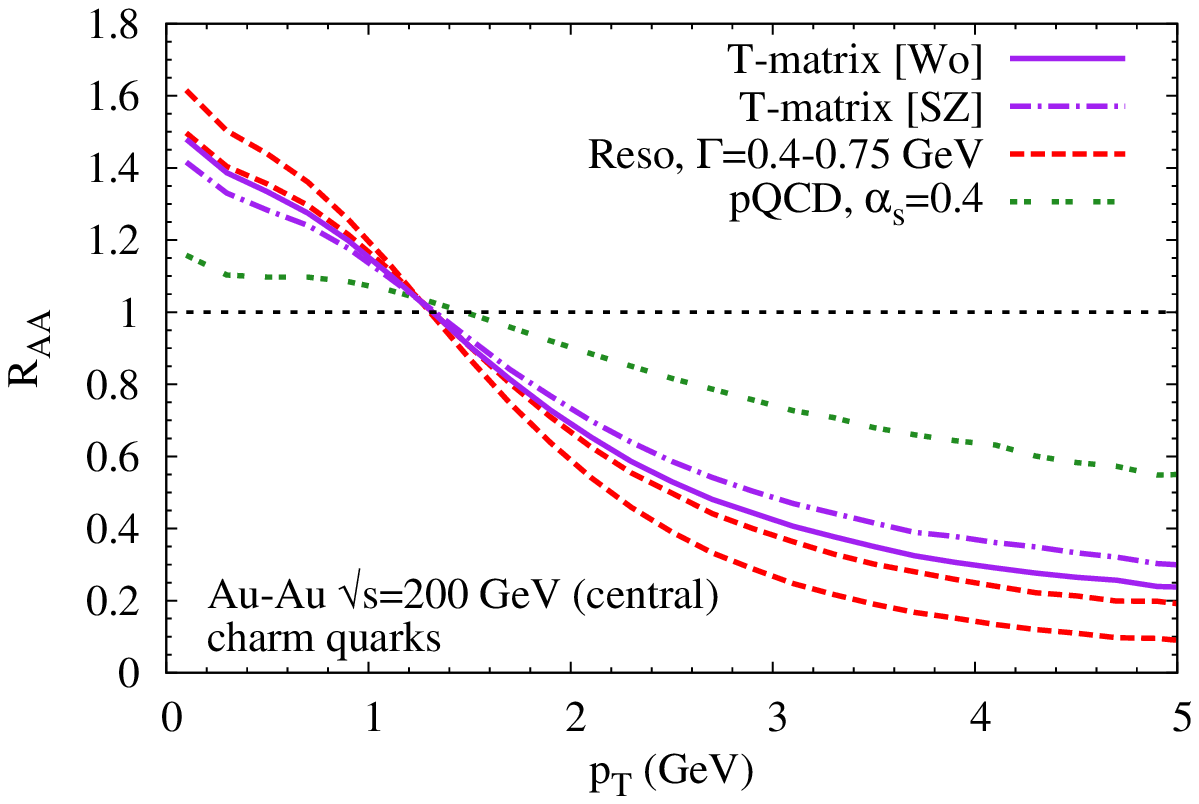}}
\centerline{\includegraphics[width=0.9\linewidth]{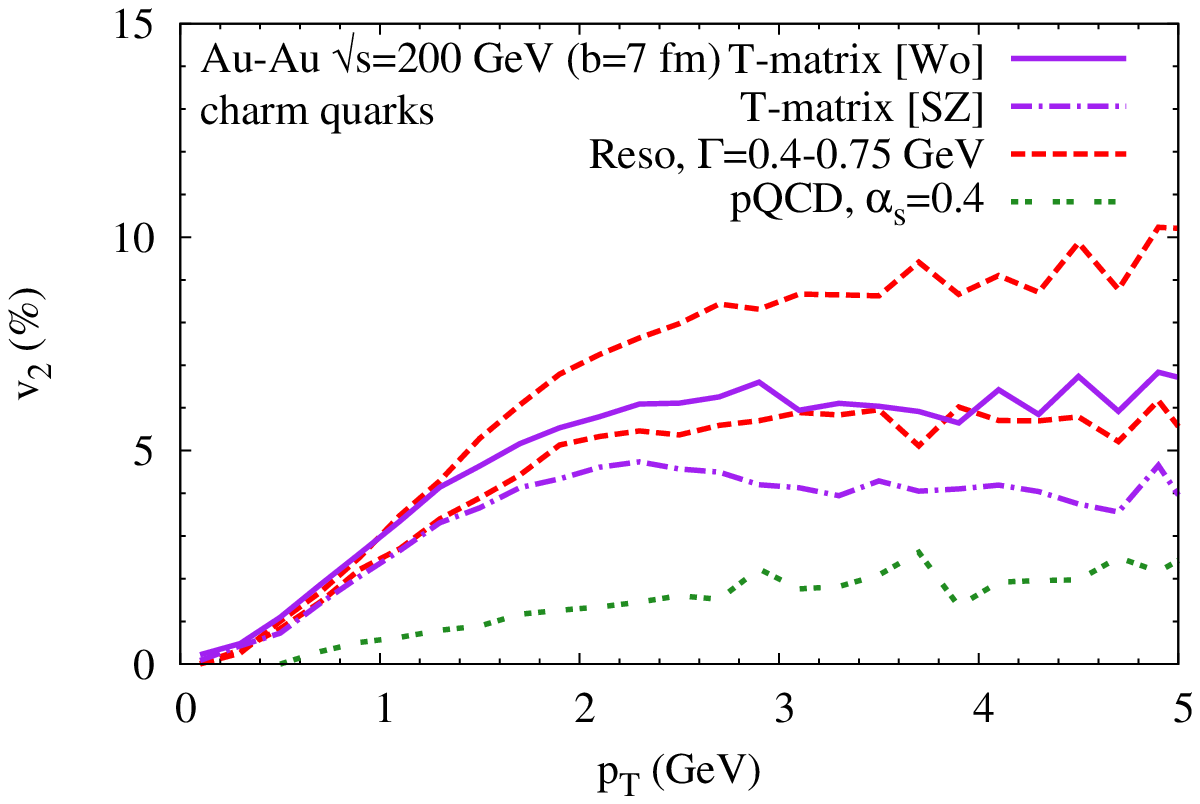}}
\caption{(Color Online) Nuclear modification factor, $R_{AA}$, for
  central (upper panel) and elliptic flow, $v_2$, for semicentral (lower
  panel) $200 \; A\text{GeV}$ Au-Au collisions for charm quarks from the
  Langevin simulation, using the $T$-matrix results for the transport
  coefficients employing the parameterizations [Wo] and [SZ] for the
  lQCD potentials, compared to a calculation based on pQCD and the
  resonance-model interactions in~\cite{vanHees:2005wb}}
\label{fig.cquark-raa-v2}
\end{figure}

\begin{figure}
\centerline{\includegraphics[width=0.9\linewidth]{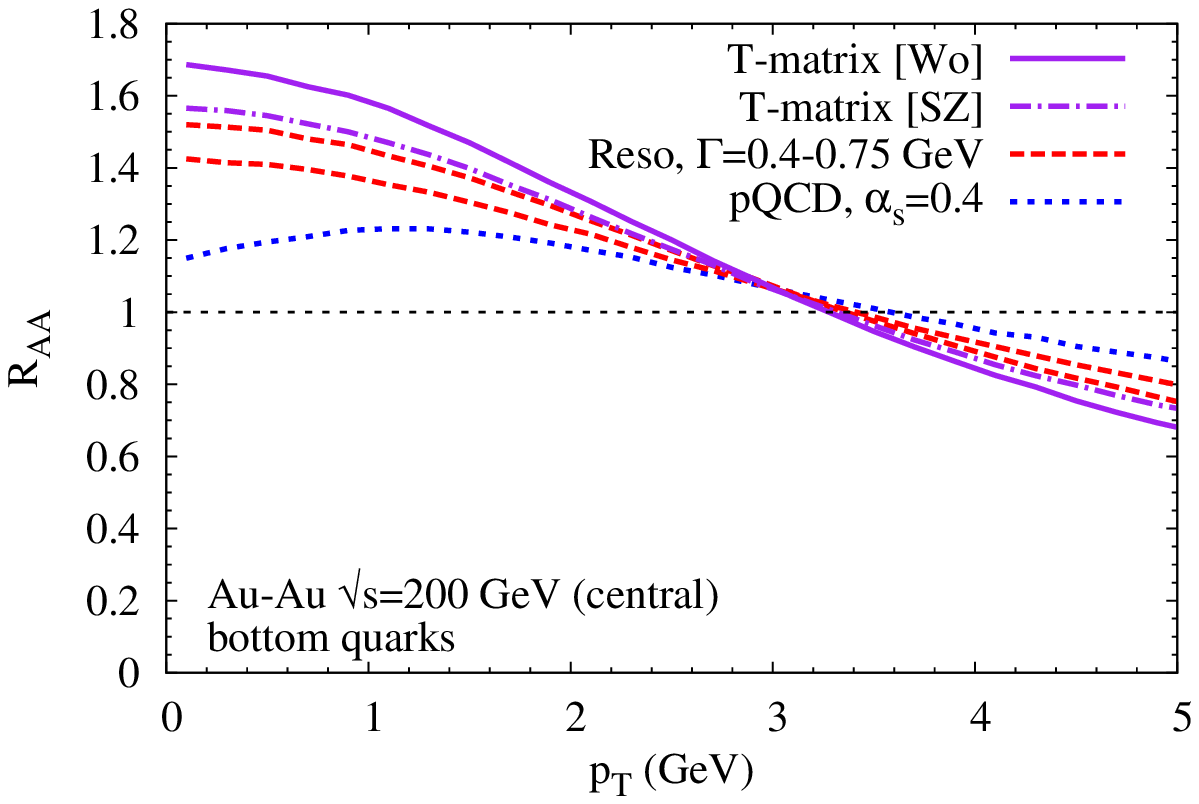}}
\centerline{\includegraphics[width=0.9\linewidth]{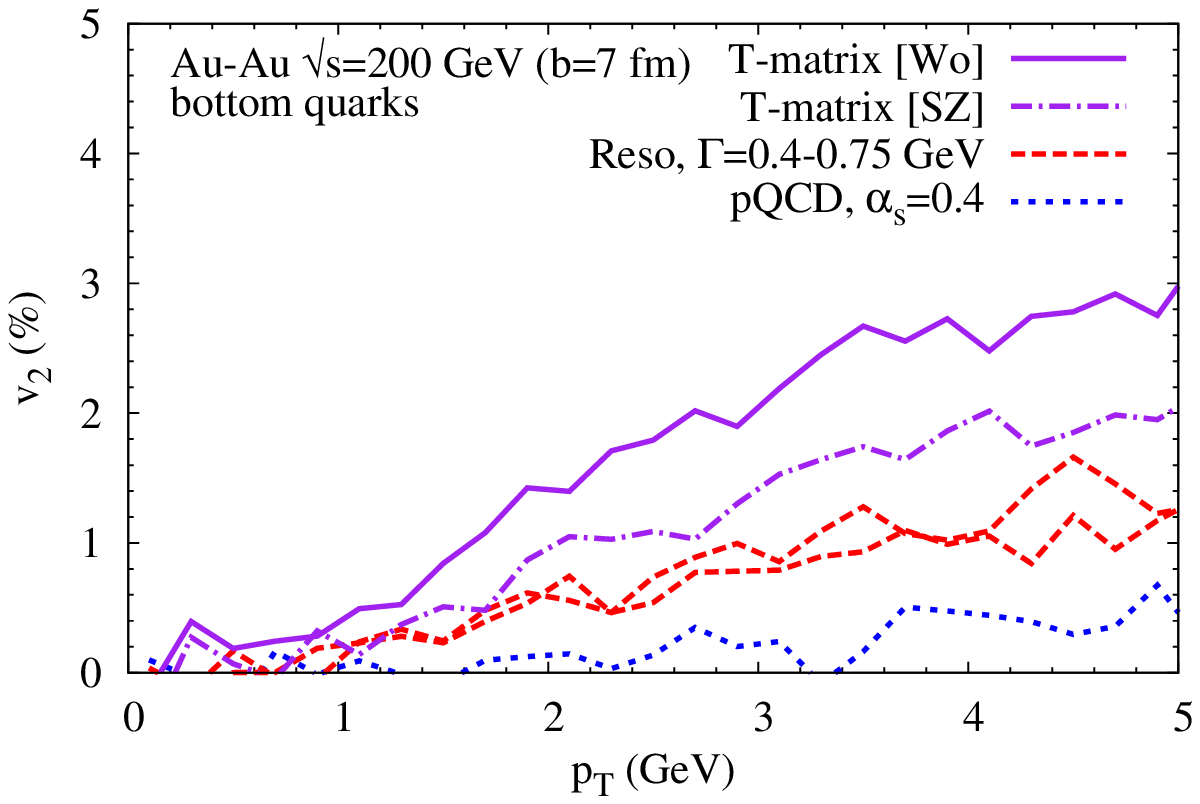}}
\caption{(Color online) The same as Fig.~\ref{fig.cquark-raa-v2} but for
  bottom quarks}
\label{fig.bquark-raa-v2}
\end{figure}

In Figs.~\ref{fig.cquark-raa-v2} and \ref{fig.bquark-raa-v2} we show the
nuclear modification factor, $R_{AA}$, defined by
$R_{AA}=P_Q(t_{\text{fin}},p_t)/P_Q(0,p_t)$ (where $t_{\text{fin}}$
denotes the time at the end of the mixed QGP-hadronic phase) in central
and the elliptic flow,
\begin{equation}
v_2=\erw{(p_x^2-p_y^2)/p_t^2},
\end{equation}
in semicentral $200\;A\text{GeV}$ Au-Au collisions. We compare the
results from the $T$-matrix model with the parameterizations by [Wo] and
[SZ] of the lQCD potentials with the those using pQCD or the
resonance-model interactions of Ref.~\cite{vanHees:2005wb}. While for
charm quarks for the [Wo] potential the result for $R_{AA}$ is
comparable to the upper end of the uncertainty band of the
resonance-model calculation, the $v_2$ is slightly enhanced at low
$p_t$. The reason for this behavior is the decrease of the transport
coefficients with increasing $T$: While the suppression of the $p_t$
spectra at high $p_t$ is due to the evolution along the whole history of
the fireball, leading to comparable effects at the end of the mixed
phase, the anisotropic flow is mostly developed at the later stages and
thus can be transferred to the heavy quarks at the end of the evolution
efficiently, when the drag coefficient become larger due to the
dynamical formation of the resonance states. The $T$-matrix result with
the somewhat less attractive [SZ] potential leads to the usual ordering
of the coefficients (increasing with increasing temperature) and thus
shows weaker effects for both the $R_{AA}$ and $v_2$ than the result of
the resonance model. For $b$ quarks the $T$-matrix calculations yield
larger medium modifications of the $p_t$ spectra than the resonance
model which is due to the mass effect, leading to stronger binding
effects for the resonances in the $T$-matrix calculation. As to be
expected, the effects of pQCD-based transport coefficients on the HQ
spectra for both charm and bottom quarks is much weaker than the
non-perturbative ones via the resonance-scattering mechanism.

\section{Single-electron observables at RHIC}
\label{sec-observables}

The last step toward a comparison of the above described mo\-del for HQ
diffusion in the QGP with the single-electron $p_t$ data from RHIC is
the hadronization of the HQ spectra to $D$- and $B$-mesons and their
subsequent semileptonic decay to $e^{\pm}$. Here we use the
quark-coalescence model described
in~\cite{Greco:2003mm,Greco:2003vf}. In the recent years, the
coalescence of quarks in the hot and dense medium created in heavy-ion
collisions has been shown to provide a successful ha\-dro\-ni\-za\-tion
mechanism to explain phenomena such as the scaling of hadronic
elliptic-flow parameters, $v_2$ with the number of constituent quarks,
$v_{2,h}(p_t) = n_h v_{2,q}(p_t/n_h)$, where $n_h=2(3)$ for mesons
(hadrons) denotes the number of constituent quarks contained in the
hadron, $h$, and the large $p/\pi$ ratio in Au-Au compared to $p$-$p$
collisions~\cite{Greco:2003mm,Hwa:2002tu,Fries:2003kq}. Quark
coalescence is most efficient in the low-$p_t$ regime where most $c$ and
$b$ quarks combine into $D$ and $B$ mesons, respectively. To conserve
the total HQ number, we assume that the remaining heavy quarks hadronize
via ($\delta$-func\-tion) fragmentation.

\begin{figure}
\centering
\includegraphics[width=0.9 \linewidth]{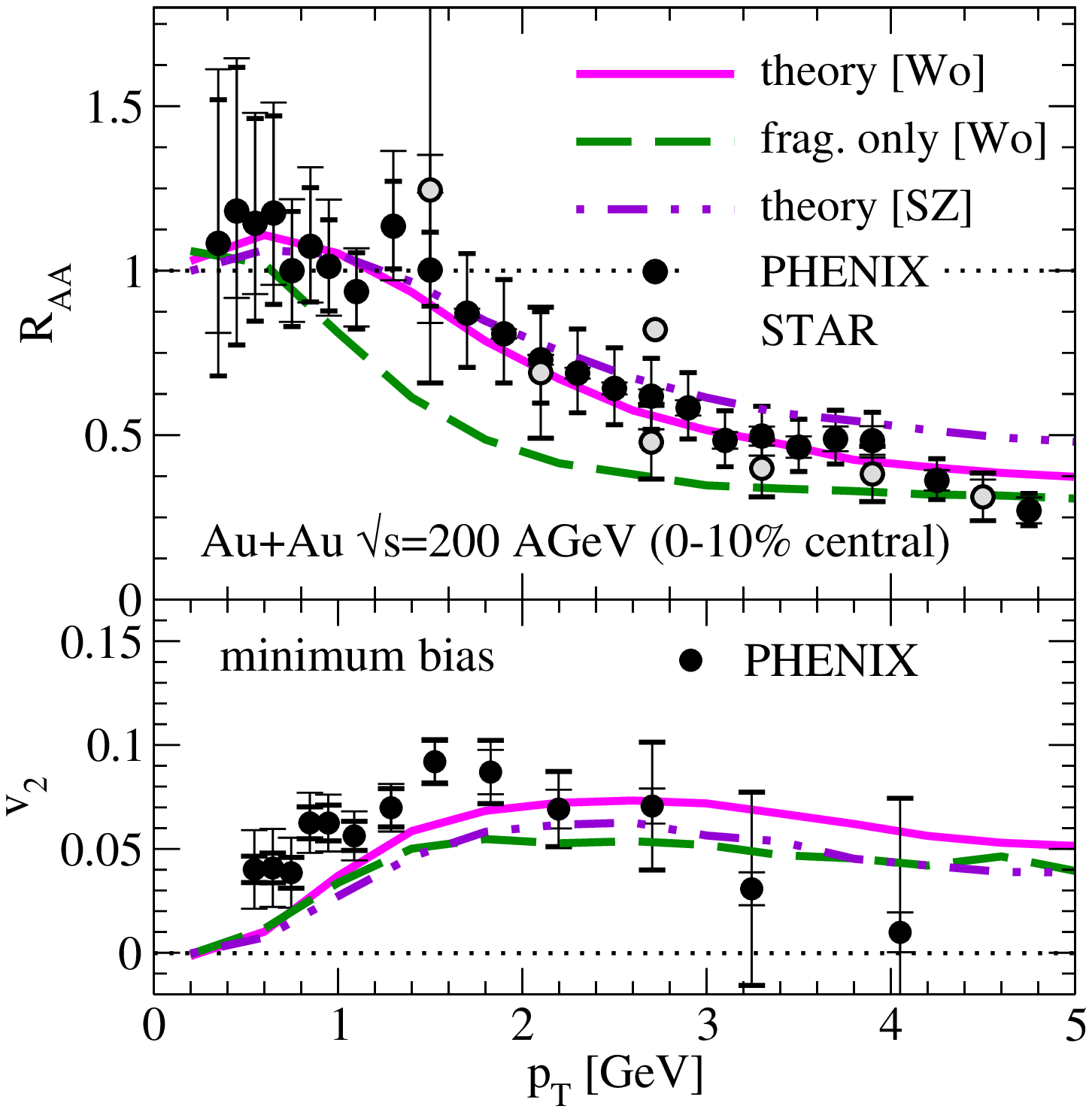}\\[0.3cm]
\includegraphics[width=0.9 \linewidth]{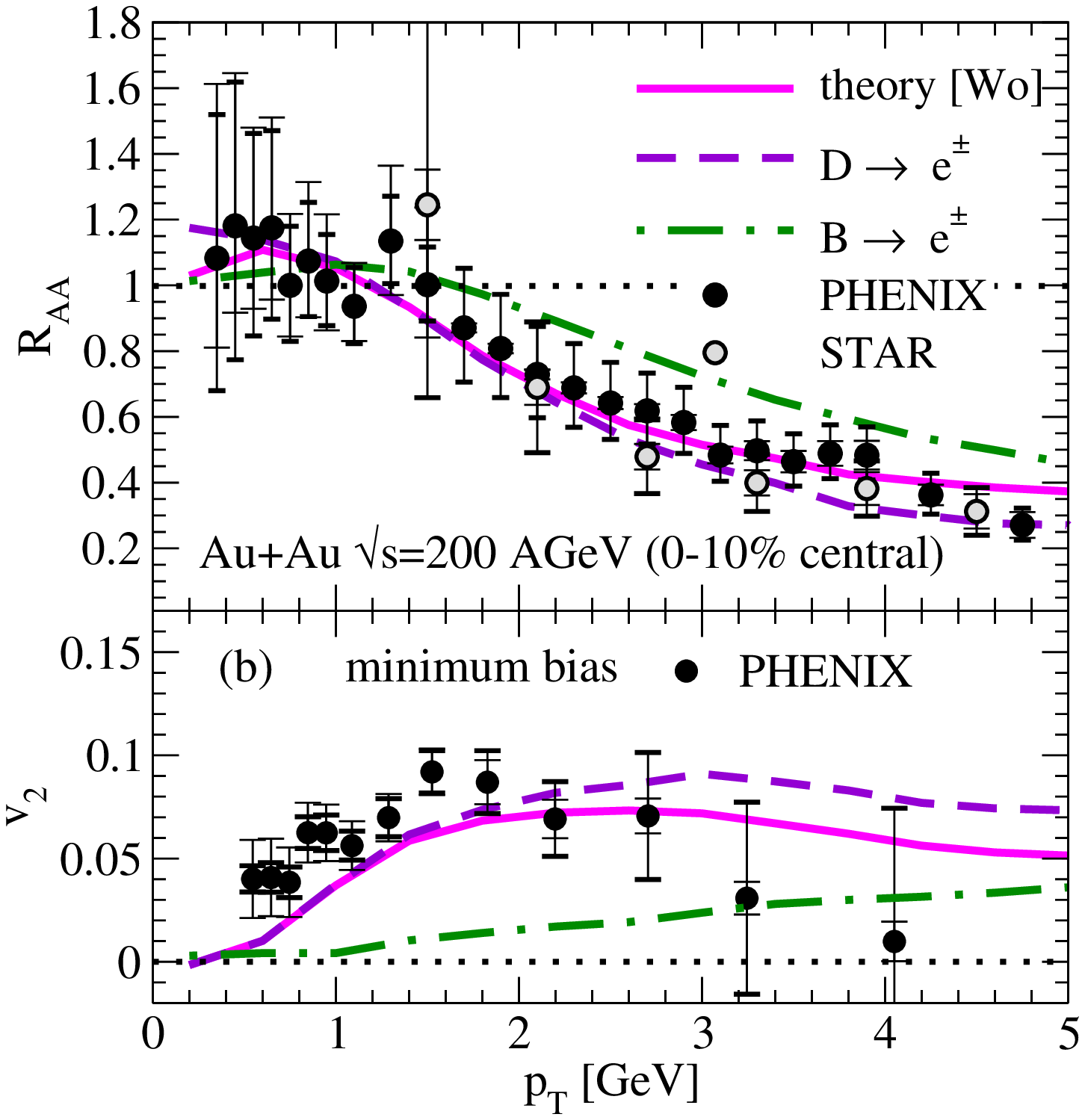}
\caption{(Color online) Upper panel: single-electron spectra from the
  $T$-matrix calculation of HQ diffusion in the QGP based on the [Wo]
  (solid line) and the [SZ] (dash-double-dotted line) parameterizations
  of the lQCD static HQ potential in comparison to data from the
  PHENIX~\cite{Abelev:2006db} and STAR~\cite{Abelev:2006db}
  collaborations in $200 \; A \text{GeV}$ Au-Au collisions at RHIC. The
  dashed line shows the result when only $\delta$-function fragmentation
  is considered for hadronization. Lower panel: $R_{AA}$ and $v_2$, as
  in the upper panel for the [Wo] parameterization for the electrons
  from $D$ (dashed line) and $B$ mesons (dash-dotted line) separately.}
\label{fig.electrons}
\end{figure}

As shown in Fig.~\ref{fig.electrons} the Langevin simulation of the HQ
diffusion based on transport coefficients from the lQCD static
potentials, followed by the combined quark-coa\-le\-scen\-ce
fragmentation description of hadronization to $D$ and $B$ mesons and
their subsequent semileptonic decay, successfully accounts
simultaneously for both the nuclear modification factor, $R_{AA}$, and
the elliptic flow, $v_2$, of single electrons in $200 \; A\text{GeV}$
Au-Au collisions~\cite{Adare:2006nq,Abelev:2006db} at RHIC. The
uncertainty due to the two different parameterizations of the potentials
by [Wo] and [SZ] is not so large. However, the deviations from other
parameterizations are bigger and will be demonstrated
elsewhere~\cite{vanHees:2008xxx}. The effects from the ``momentum kick''
of the light quarks in quark coalescence, an enhancement of both,
$R_{AA}$ and $v_2$, is important for the quite good agreement of both
observables with the data. As can be seen from the lower panel in
Fig.~\ref{fig.electrons}, within our model the effects from the mixing
of the $B$-meson decay contribution to the $e^{\pm}$ spectra becomes
visible in the region of $p_t \simeq 2.5$-$3 \; \text{GeV}$. A closer
inspection of the time evolution of the $p_t$ spectra shows that the
suppression of high-$p_t$ heavy quarks occurs mostly in the beginning of
the time evolution, while the $v_2$ is built up later at temperatures
close to $T_c$ which is to be expected since the $v_2$ of the bulk
medium is fully developed at later stages only. This effect is also
pronounced for the [Wo] parameterization of the HQ potential since in
this case due to resonance formation the transport coefficients become
largest close to $T_c$.

\section{Conclusions and outlook}
\label{sec-conclusions}

We have used static potentials from finite-temperature lQCD calculations
within a Brueckner-type many-body calculation, complemented by pQCD
HQ-gluon elastic-scattering matrix elements, to assess drag and
diffusion coefficients for $c$ and $b$ quarks in the QGP in a
principally parameter free approach, however plagued with large
uncertainties in the determination of the relevant potential from
lattice data. The diffusion of heavy quarks in the QGP is calculated
with a Langevin simulation. The medium is parameterized as an expanding
thermal fireball (including an\-iso\-tro\-pic flow for semicentral
heavy-ion collisions) with an equation of state of a massless gas of
gluons and $N_f=2.5$ light-quark flavors. To confront this model with
data on non-photonic single electrons in $200 \; A\text{GeV}$ Au-Au
collisions at RHIC, we have used a combined coalescence-fragmentation
model to hadronize the heavy quarks to $D$ and $B$ mesons which
subsequently decay semi-leptonically. The resulting $p_t$ spectra agree
with recent data on the nuclear modification factor, $R_{AA}$ and
elliptic flow, $v_2$ quite well.

In a schematic estimate from the evaluated drag and diffusion
coefficients leading to this results, based on kinetic theory using
either pQCD for a weakly coupled plas\-ma
\cite{Israel:1970,Danielewicz:1984ww} or the strong-coupling limit
applying AdS/CFT
correspondence~\cite{Herzog:2006gh,CasalderreySolana:2006rq}, we find
values for the space-dif\-fu\-sion coefficient $2 \pi T D_s=4$-$6$ and a
viscosity to entropy-density ratio of $\eta/s=2$-$5/(4 \pi)$ (to be
compared to the conjectured AdS/CFT bound $(\eta/s)_{\mathrm{min}}=1/(4
\pi)$), indicating a strong\-ly coupled (liquid like) quark-gluon plasma
close to the phase transition~\cite{Rapp:2008qc}.

In future works detailed studies of the uncertainties in the potential
approach is necessary, in particular about the question whether a
static-potential approach is justified and which in-medium potential
(i.e., free or internal potential energy or combinations thereof) should
be used to describe the interactions of heavy quarks within the
sQGP. First steps in this direction have been made
in~\cite{Brambilla:2008cx} in an effective-field theory approach for
non-relativistic in-medium QCD-bound states. Also an inclusion of
inelastic processes like gluon bremsstrahlung which should become
effective at higher $p_t$ is mandatory for a complete picture of the
in-medium behavior of heavy quarks~\cite{Vitev:2007jj}.

Another step, which is quite natural, given that with\-in our model the
underlying microscopic effect of the strong coupling of the heavy quarks
to the medium is the formation of resonance states close to $T_c$, is to
substitute the quark-coalescence model with a transport-model based
resonance-recombination model for hadronization in the
QGP~\cite{Ravagli:2007xx} which obeys the conservation laws for energy
and momentum as well as the second law of thermodynamics. This model
shares with quark-coalescence hadronization the phenomenologically
successful feature of the scaling of $v_2$ with the hadrons' number of
constituent quarks. Additionally it leads to scaling with the transverse
kinetic energy well, provided the parton distribution and flow is
calculated with a dynamically consistent scheme as the Langevin
simulation discussed in the present paper~\cite{Ravagli:2008rt}.

\begin{acknowledgement}
  This work has been supported by the U.S. National Science foundation
  under CAREER grant PHY-0449489 (HvH,RR), and by the Ministerio de
  Educaci{\'o}n y Ciencia under grant AYA 2005-08013-C03-02 (MM).
\end{acknowledgement}


\end{document}